# Where to find the mind: Identifying the scale of cognitive dynamics

Luke Conlin, Department of Curriculum & Instruction, Benjamin Building, College Park, MD 20742 USA, luke.conlin@gmail.com
Ayush Gupta, Department of Physics, Toll Building, College Park, MD 20742 USA, ayush@umd.edu
David Hammer, Departments of Physics and Curriculum & Instruction, , Toll Building, College Park, MD 20742 USA, davidham@umd.edu

**Abstract:** There are ongoing divisions in the learning sciences between perspectives that treat cognition as occurring within individual minds and those that treat it as irreducibly distributed or situated in material and social contexts. We contend that accounts of individual minds as complex systems are theoretically continuous with distributed and situated cognition. On this view, the difference is a matter of the scale of the dynamics of interest, and the choice of scale can be informed by data. In this paper, we propose heuristics for empirically determining the scale of the relevant cognitive dynamics. We illustrate these heuristics in two contrasting cases, one in which the evidence supports attributing cognition to a group of students and one in which the evidence supports attributing cognition to an individual.

## Introduction

Researchers have been divided on how to answer the question "Where is the mind?" (Cobb, 1994) The "cognitivist" perspective takes the individual as the unit of theoretical and empirical focus, under the assumption that cognition is something that only brains do. While cognitivists do not ignore the social and material contexts in which cognition takes place, they analyze cognition by decomposing it into the information processing of individuals (Anderson, Reder, & Simon, 1997). That stance contrasts with situated and distributed views, which contend that cognitive processes cannot be neatly attributed to contributions of individuals when they interact with each other and their environment. Arguments for situated cognition often cite the example of a dieter solving the problem of taking three quarters of half a cup of cottage cheese by pouring out half a cup of cottage cheese, cutting it into fours, and removing one quarter (Lave, Murtaugh, & De La Rocha, 1984). A paradigmatic example of distributed cognition describes the information processing within an airplane cockpit to show that it is the cockpit as a whole—the people and instruments together as a unit—that "remembers" the safe landing speed of the airplane (Hutchins, 1995). Each challenges the attribution of cognition to an individual mind by arguing that cognition is irreducibly situated in a material and social context. The dieter's knowledge of finding ¾ of ½ a cup is inseparable from the physical materials they use to solve the problem. Likewise, no one part of the cockpit "knows" the ever-changing safe landing speed of the airplane.

There have been a number of arguments regarding the relationship between these perspectives. Cobb (1994) took them to be complementary, with one in the background of the other; Sfard (1998) also considered them complementary but "incommensurable"; Greeno (1997) took them as competing alternatives. In this article, we review an argument for theoretical continuity among these perspectives, by which they represent different scales of a complex system, rather than as fundamentally different approaches. Rather than consider the choice of perspective a matter of *a priori* theoretical commitments, we propose treating that choice as empirical. We offer heuristics for matching the choice of unit of analysis to the dynamics evident in the data, arguing that the cognitive unit can be modeled at the grain size(s) at which there are observable stabilities and coherent dynamics in cognitive activity. We then illustrate the use of these heuristics with two example analyses demonstrating empirical evidence for group and individual cognitive units, respectively.

## Manifold resources

Minsky's (1988) is probably the most widely known account of mind as comprised of manifold cognitive and metacognitive parts. He described a "society of mind" made up of a very large number of "agents." One agent he proposed was *More*. When *More* was active it was comparing two amounts and deciding which was greater. *More* was a society itself, made up of agents that corresponded to different ways of comparing different sorts of amounts, and the agents that made up more were made up of other agents. The model presented a complex system of interactions and structures at a wide range of scales. That work influenced many others' accounts, including diSessa's "knowledge in pieces" (1988) and Dennett's (1991) multiple drafts model of consciousness; it resembles and may have been influenced by "schemas" (Rumelhart, 1980).

We refer to "resources" as a generic term for the parts in such models of mind (Hammer, Elby, Scherr, & Redish, 2005). On this view, what we experience as "states of mind" correspond to stable patterns of resource activations, stability that might be only local and contextual or over extended timescales; a stable pattern of resource activations may become a resource in itself, much like Minsky's *More*. But a stable pattern of activated fine-grained cognitive elements is not necessarily constrained to an individual mind: the pattern and its stability might extend over multiple individuals and artifacts. In this sense, the unit of "cognition" – the pattern of resources and its stability – might at times be individual or be distributed. Our claim is that empirical determination of the extent and locus of stability of such cognitive coherence should guide our sense of the unit of cognition, not *a priori* theoretical commitments.

Consistent with the views of cognition as situated and distributed, there are behaviors and cognitive dynamics that can emerge out of the interaction of multiple individuals and artifacts. Great strides have been made in the last quarter century in modeling how complex behaviors can emerge from interacting agents following surprisingly simple sets of rules. Paradigmatic examples include flocking birds (Kennedy & Eberhart, 1995), synchronized flashes of fireflies (Mirollo & Strogatz, 1990), and foraging of ant colonies (Dorigo & Stützle, 2004). Thelen and Smith (1994; Smith, 2005) have argued for complex systems-based models of cognitive development in individuals, again taking a mind to be made up of many interacting parts in dynamic interaction with each other.

Meanwhile, social science has advanced a variety of accounts of interactive dynamics, such as in conversations tending to abide by simple rules (Grice, 1989) and in shared interpretive frameworks (Goffman, 1974) that are maintained explicitly and implicitly by attending to and responding to verbal, facial and behavioral metamessages (Bateson, 1985; Tannen, 1993). These too may be understood as complex systems, at this larger scale. We suggest that these can be seen as similar dynamics at different scales, that a "society of mind" model of individual cognition is theoretical continuous with a "mind of society" model of social cognition. Thus one can model the dynamics of cognition via the synchronized activity of many cognitive resources, whether those resources are within an individual mind or distributed across minds and materials.

## What to look for: Heuristics for Data-Based Choice of Unit of Cognition

On this view, rather than as *a priori* commitment, the scale of the relevant dynamics may be determined by the evidence at hand. Our core purpose here is to propose four heuristics for making that determination: *clustering*, *persistence*, *resistance*, and *transition*. We extracted these from a range of recent work dealing with a variety of dynamics relevant to cognition including conceptions, behaviors, framing, and epistemologies (Conlin, Gupta, Scherr, & Hammer; Frank, Kanim, & Gomez, 2008; Scherr & Hammer, 2009). The heuristics are not a coding scheme, but rather a guide for empirically grounding the choice of the unit of analysis. In what follows we describe the heuristics and then apply them to two specific examples to show how they can do work in finding an appropriate unit of analysis for modeling cognition.

### Clustering

The basic idea of clustering is that a coherent cognitive state can be characterized by the 'hanging together' or simultaneous activation of multiple elements in the same domain or across domains. For example, how we recognize someone as greeting us is by the clustering of behavioral and speech patterns such as a hand extended for hand-shake, a smile on the face, an elevated vocal pitch, and the utterance of socially-accepted greeting words such as "Hello." Together, these behaviors help us interpret the activity as a "greeting." Scherr & Hammer (2009) found clusters of various kinds of behaviors (e.g., facial expressions, gestures, pitch, posture) that tended to occur synchronously within and across students working on physics worksheets in groups of four.

### Persistence

If clusters are to be suggestive of cognitive stabilities then they need to be more than just fleeting happenstance; they should persist over time. In the case of a hand-shake the behaviors might persist for a few seconds, while in other cases clusters could remain stable over much larger time scale. For example, while watching a thriller movie, behaviors such as stiff posture, intense gaze on the screen, and low perceptivity of the environment could form a cluster that persists over several minutes. In (Scherr & Hammer, 2009) stable clusters of behaviors were found to last from as little as a few seconds up to several uninterrupted minutes.

### Resistance to Change

In general, the stability of the state of a system is inferred in its ability to sustain itself in the face of perturbations. In certain moments, minor changes in the makeup of the cluster (whether the unit is the individual or the group) could challenge the coherence of physical and cognitive activity. Still, the clusters characterizing the cognitive coherence may be so stable that they can persist despite the perturbation/disturbance. Think of a greeting that lacks an important element, say, when one person extends a hand but the other person does not shake it. This can be a palpable challenge to the "greeting" frame, but it does not always result in a breakdown of the rest of the greeting activities. Fine-grained analysis of the phenomenon can illuminate how particular clusters react to naturally occurring perturbations, providing evidence on the stability and transition dynamics of clusters. More importantly, such analysis can provide empirical insight into whether the cluster (of resources, behaviors, etc.) in that moment is stabilized by an individual or a group.

Transitions

Sudden transitions in behavior help mark the spatial and temporal boundaries of the stabilities. If there are multiple stable states in the cognitive dynamics of an individual or a group, then we should be able to see distinct transitions amongst the stable states. For this reason it is often easier to notice changes in clusters than to notice the stable clusters themselves—when something suddenly changes it highlights what was just a local stability. When a greeting is over, all of the behaviors that are characteristic of greetings end at about the same time, possibly turning into another cluster corresponding to, say, a conversation.

For an empirical example, the behavioral clusters of Scherr and Hammer (2009) were found only after noticing distinct transitions in behavior. That there were noticeable transitions led the researchers to find and clarify what the transitions were *from* and *to*, which upon further analysis were found to be coherent clusters of behaviors. The behavioral transitions are easily and reliably identified: before discussion three independent coders agreed on the timing of 90% behavioral transitions (to within five seconds). Transitions are often the result of "bids", i.e. minor changes in the clusters of behaviors. Bids that do not precipitate a transition to a new behavior cluster are evidence of the stability of the cluster. Bids that lead to a marked transition between stable clusters are also evidence of the coherence of those clusters.

In the examples that follow we will report on the clustering, resistance, and transitions of phenomena at the individual level and at the group level. We attend to the students' physical behavior (gesture, gaze, posture, etc.) as well as the substance of their discourse in applying these heuristics.

## Data Analysis

Our data is from an introductory algebra-based introductory physics course at a large public university. The students are mostly third-year life sciences majors. We draw our observations from video taken during tutorial sessions – 50-minute sessions, facilitated by teaching assistants, in which the students work in groups of four on ungraded guided-inquiry worksheets focusing on conceptual and epistemological development in physics.

First, we present an example of group-level dynamics in behavior and *epistemological framing*, that is a sense of "what is it that is going on here" with respect to knowledge (Hammer, *et al*, 2005). In the process we will show how the evidence of clustering, persistence over time, resistance to change, and spontaneous transitions all support this account in terms of group-level stabilities and dynamics. We then present a second example that supports an individual unit of analysis but not group-level cognition, in order to show a contrasting case of the scale of the unit of analysis.

More typically, we expect, the relevant scale may shift over the course of an episode, and we expect that analyzing those dynamics will be a fruitful direction for further analyses. For the present, we focus on articulating heuristics for identifying the scale of the dynamics, and for that purpose we have chosen "cleaner" examples.

## The Group as a Unit of Analysis

The first example[i] comes from a Newton's Third Law tutorial, where four students (Amanda, Bridget, Camille, and Dianna) are working together on a series of worksheet questions about a collision between a truck and a car. The particular question they are considering during this clip asks: "Suppose the truck's mass is 2000 kg while the car's mass is 1000 kg. And suppose the truck slows down by 5 m/s during the collision. Intuitively, how much speed does the car gain during the collision?" Camille is the first to offer a response, asking whether the car gains 5 m/s since the truck slows down by 5 m/s. This is an excerpt from the ensuing discussion:

| 1 | Amanda: | The car is half as heavy, so it'll gain twice as much |
| 2 | Camille: | Ah, shoot (laughs) |
| 3 | Dianna: | Or something, Idunno. |
| 4 | Amanda: | That's what they want us to think, but this is not the real answer |

| 5 | Bridget: | This is not the right /one/. Apparently, I think that's what they want us to say. |
|---|---|---|
| | | *…about ten seconds of silence…* |
| 6 | Dianna: | This is going…five…five meters per second, that's it's what? Acceleration or velocity? |
| 7 | Camille: | Speed. Velocity |
| 8 | Amanda: | (together with Camille) velocity. |
| 9 | Bridget: | Slows down by…?? |
| 10 | Dianna: | Velocity? |
| 11 | Camille: | Mm hmm. (pause) So the car gains ten meters per second? |
| 12 | Bridget: | I guess. |
| 13 | Dianna: | Didn't he say something about how like…something in class, like…if something's touched, the velocity, or something was changed…what was he talking about in class? Something? |
| | | *…about ten seconds of silence…* |

Camille's first intuition differed from Amanda's and Bridget's. Camille says, "the truck slows down by 5 m/s, the car speeds up by 5 m/s" but Amanda says that it would be 10 m/s because "the car is half as heavy." The discussion, however, quickly dies out and the students continue to engage in the school routine of completing the worksheet (writing best-guess answers, and occasionally seeking confirmation of vocabulary terms), until Dianna overhears the word "intuitively" from a different group's conversation, and expresses her own frustration:

| 14 | Dianna: | I hate that word, "intuitively." |
|---|---|---|
| 15 | Camille: | See intuitively, I would think that it'd slow down, I mean speed up five meters per second. |
| 16 | Bridget: | Isn't the car…do they mean /once they hit/? |
| 17 | Amanda: | Because it is slowing down. |
| 18 | Camille: | If the if the truck if the truck if the truck, suppose the truck slows down by five meters in the collision, so if the truck is slowing down, then /I guess/ the car has to be speeding up. |
| 19 | Dianna: | The car's not moving |
| 20 | Amanda: | Yeah, it's not moving |
| 21 | Bridget: | But I think…does that mean- |
| 22 | Camille: | No, but I'm saying it says how much does does the car gain, cause it, yeah… |
| 23 | Bridget: | cause like it goes psh (gestures collision with hands) |

A vigorous discussion follows Dianne's statement as the students now focus on their own reasoning, not the worksheet; their voices and behaviors are animated as they try to communicate their ideas with gestures and explanations; and they pose original questions, such as what if the car and truck were equally heavy (not shown in transcript). During this strip of discourse, they clarify the details of the collision and the resulting speeds.

In what follows, we apply our heuristics to empirical observation of the students' behaviors and reasoning in these two segments to argue for the *group, rather than individual students* as the unit that determines the stability and dynamics of the behavioral and reasoning patterns.

## Clustering of Behavior and Reasoning: Group as Unit

As students go through the tutorial, multiple elements of their behaviors – gaze, body posture, extent of gestures, and tone of voice – are coordinated or clustered in particular patterns. Scherr and Hammer (2009) identified four stable clusters of behaviors exhibited by groups in tutorial, and these clusters account for most of the groups' time spent in tutorial (Conlin, Gupta, R. E Scherr, & D. Hammer, 2007) We will only mention the two clusters that are relevant here, *blue* and *green*.

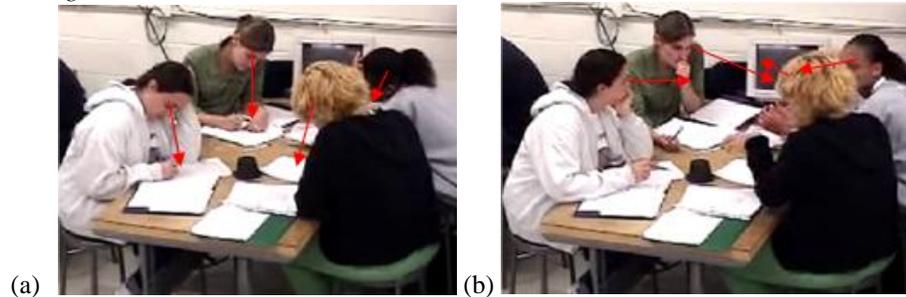

(a) (b)

Figure 1. (a) The students exhibiting the "blue" cluster of behaviors, and (b) the students exhibiting the "green" cluster.

In the first segment of the episode, the group is exhibiting the *blue* behavioral cluster (see Figure 1a) -characterized by downward gaze, hands writing or resting on the table, scant and mostly deictic gesturing, low or flat voices, and intermittent speech with no overlap. In the second segment, the students transition *as a group* to the *green* behavioral cluster (Figure1b), in which they tend to make eye contact, sit up, gesture, and speak in animated voices with overlapping speech. The arrows in Figure 1 show how the gaze of the students is coordinated – and the locus of their gaze shifts in synchronous patterns making it meaningful to talk about the *group* as gazing towards the worksheet during the earlier segment and towards one another during the second segment.

These behavioral clusters also often indicate locally stable ways in which the students were framing the activity, either as *completing the worksheet* or as *having a discussion* (Scherr & Hammer, 2009). At first, their speech is in the service of finding the answer to the worksheet question. Amanda, Dianne, and Bridget all but say this explicitly by suggesting that Camille write down an answer that "they want us to think" rather than one that makes sense to her. Furthermore, their answers are brief and only weakly supported with justification. Dianna tries to settle the vocabulary ("five meters per second, that's its what, acceleration or velocity?") and tries to remember the information the professor gave in class about collisions. The content of their speech taken in conjunction with their behaviors suggests that the group is initially framing the activity as *completing the worksheet*. After Dianna expresses her frustration with the word "intuitively," the group starts to clarify and piece together the mechanism of what happens during a collision, rather than simply comparing the speed before and after. Their speech and behaviors are clustered around *having a discussion* about a physical phenomenon: Bridget, Amanda, and Dianne try to clarify the conditions of the collision; Camille reasons about the reactions of the truck and car as related, "if the truck *is slowing down*, then /I guess/ the car *has to be speeding up*" (emphasis added); Bridget's gestures try to simulate the motion of the car.

## Transition in Behavior and Reasoning: Group as Unit

There is a sharp transition in behavior at both the individual and the group levels that starts with Dianna's commentary that she "hates that word, 'intuitively.'" Her affective stance shook the group away from the worksheet, opening up the space for a more personal discussion. Everyone looks up to pay attention to Dianna – a sharing among friends. To Camille, this provided the space to clarify her original intuition, an action she did not take in the first part of the episode when the focus was on figuring out what "they want us to think." Camille's emphasis on the word 'intuitively' and tone support this interpretation. Next, Camille emphasizes the compensation of speeds between the truck and the car; Amanda gestures the speeding up of the car after the collision. The emotional outburst of Dianna cued them to think about and communicate their personal opinions, leading to a cascade of utterances towards mechanistic pursuit of the phenomenon. The coordinated nature of the transition in behavior (from *blue* cluster to the *green* cluster) and framing (from *doing the worksheet* to *having a discussion*) makes it appropriate to think about the *group* as transitioning rather than four individuals separately making these transitions. This supports the theoretical coupling of framing and behavior, since behaviors act as the metamessages that comprise and maintain the group's framing (Bateson, 1985).

Resistance of Behavior and Reasoning: Group as Unit
*Bids to switch frames* – instances in which one students' behaviors and/or reasoning stand in contrast to that of the established coherent frame (Scherr & Hammer, 2009) – can provide valuable insight into the dynamics and appropriate scale for a cognitive unit. In this episode, while the group is exhibiting the blue behavioral cluster engaged in *completing the worksheet*, Dianna makes a bid to have a discussion (turn 13). This is evident through her behavior, since she sits up, speaks more loudly, and has much more pronounced prosody when asking, "Didn't they say something about how like...what was he talking about in class?" One by one, the other students briefly look up as she is talking. As she trails off, however, the other students return to their blue cluster behaviors one by one. There is no talking for 10 seconds as the group goes back to completing the worksheet. Dianna's bid to change behaviors was not taken up.

Epistemologically, Dianna's question is a bid to try to remember what the professor said in class, a departure from the group's engagement in finding the worksheet-directed intuitive answer to the gain in the car's speed. That the rest of the group simply turns back down to their worksheets without responding to her question sends the message that they did not see this as a productive line of pursuit. Dianne's deviation from the shared cluster of behaviors and framing could not disrupt it, rather it was Dianne who returned to that shared activity. This illustrates the stability of the cluster and its extension beyond single individuals.

Persistence of Behavior and Reasoning: Group as Unit
Additional evidence for the *group* being the appropriate unit of analysis here comes from examining what keeps the behavior or framing stable during these segments. We find that the behaviors serving as meta-messages communicating the fine-grained stances students are taking in their utterances. When a student makes animated gestures or talks loudly, it sends the message of a discussion or argument in contrast to hunching over your personal worksheet. Seeing one member clarifying an idea might prompt a second student to add to it, which supports and encourages the effort of the first creating feedback loops that help in the persistence of the activity. The moment-to-moment behaviors and utterances establish a shared understanding of the nature of the activity – that is self-sustaining. Before the transition, individual behaviors such as hunching over the worksheet, exchanging short vocabulary or quick questions in low voices, reinforces the *group's* shared sense of the activity of filling the worksheet; and later, after the transition, their pursuit of a mechanistic explanation of what happens in the collision is built out of bits of contribution from the different members. In either segment, the persistence of behaviors and reasoning is a group endeavor, even though individuals are enacting these behaviors and utterances.

## The Individual as a Unit of Analysis

The second example[ii] comes from a tutorial on shadows and light. In this tutorial, students try to build a model of light by exploring the patterns and motions of light created by a light bulb and an aperture. The question being considered by the group during the clip is, "Why does the light on the screen moves to the left when the bulb is moved to the right?" Veronica and Jan have a disagreement over the mechanism of light shining through the aperture. Here is a piece of the conversation:

| Jan: | All the rays are going like this. So, it's kind of like polarizing it. |
|---|---|
| Veronica: | Mmm, not really. |
| Jan: | You sure? |
| Veronica: | It's just, well, it's just, guys, you're making it- you're trying to make it too difficult. It's just, the light goes out. It only goes through that one circle. So, obviously, if it is down here, and I'm looking through that circle. Look, you're sitting down here. You're looking at this big cardboard. You're looking up through that little circle. All you're going to see is what's up there. It's a direct line. |
| Jan: | Look, I see what you are saying, alright? But, I'm just trying to make it like physics-, physics-oriented. |
| Veronica: | It *is* physics-oriented, that's just the way it is. |
| Jan: | Okay. |

### Clustering of Behavior and Reasoning: Individuals as Unit

Throughout the conversation presented as well as before and after it, Veronica is sitting up, looking at the other group members, speaking in a loud, animated tone, and gesturing (Fig 2). Jan, who is sitting across the table from Veronica, for large parts of the tutorial, is hunched over, and drawing on the tutorial worksheet. During the moments of the conversation above, she does look up at Veronica and speak in a clear and loud voice, but these were fleeting moments. The larger pattern was of students conducting themselves differently. This lack of synchronicity and coordination among the behaviors suggests a clustering at the individual level in this case, not at the group level.

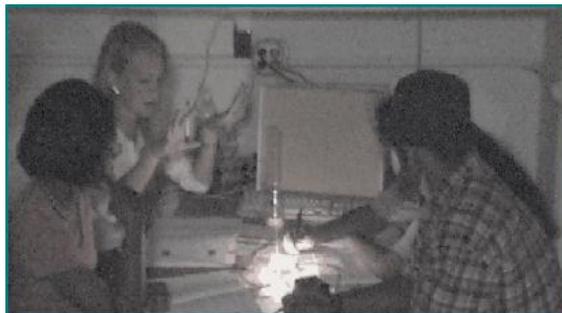

Figure 2. Veronica is gesturing as Jan is writing on her worksheet.

Again, the contrasting behaviors indicate that Jan and Veronica are framing the activity differently, also evidenced by their very different epistemological stances. Veronica is appealing to her intuitive sense of mechanism of how light travels, while Jan tries to make what Veronica is saying more "physics-oriented," by using technical terminology. Veronica objects with a different take on what it means to be physics-oriented, namely that it is "just the way it is." For Veronica, the activity is about making sense, while Jan is trying to get the formal answer.

### Persistence and Resistance of Behaviors and Reasoning: Individuals as Unit

Both Veronica and Jan activate a stable set of epistemological resources throughout this clip. In fact, this stability persists on much larger timescales than one tutorial period. Lising & Elby (2005) analyze Jan's work throughout the course and argue that her epistemological stance, in which there is a barrier between formal and everyday reasoning, is relatively stable and robust throughout the semester.

The persistence of each student in their epistemological framing is stable in spite of direct challenges from each other. Veronica expresses impatience with Jan by bluntly correcting her with, "Mmmm, not really." She criticizes Jan for "trying to make it to difficult," and proceeds to explain the phenomenon with common-sense reasoning. Jan contents that she is just trying to make it "physics-oriented," with the implication that Veronica's common-sense explanation is not physics-oriented. Veronica defends herself from this challenge: "It *is* physics-oriented. It's just the way it is." When Jan responds with, "Okay," the group goes silent. It appears that they just agree to disagree, persisting in their epistemological framing despite the explicit bids made to change them.

### Transition in Behaviors and Reasoning: Individuals as Unit

Just as there is no coherent clustering at the group level, there are no clear transitions at the group level between clusters. This seems to be the exception rather than the norm, since most of the time groups spend in tutorial is in one of four behavioral clusters separated by sharp transitions (Scherr & Hammer, 2009). This counts as evidence *against* the group being the unit of cognitive analysis during this clip. This lack of group transitions, taken in conjunction with the other heuristics above, constitutes a strong empirical case for taking the individual as the unit of analysis here.

## Conclusions

Our argument is two-tiered. On one tier, we analyze the behavior and discourse within two episodes of student group work in introductory physics tutorials. We found evidence for stabilities and dynamics of cognition at the group level and the individual level, using four empirical heuristics for identifying the scale of the cognitive unit. In the first example we demonstrated that the clustering, persistence, resistance, and transition reflect dynamics at the level of the group as a whole. It is the collective group frame that enables them to reason about the mechanisms involved in a collision in a more sophisticated way. In the second example we demonstrated how clustering, persistence, and resistance resided not with the group but rather on the level of individual students' epistemological stances and their conceptual understanding of how light travels. Veronica and Jan held different views not only of light but also of physics, and these views remained stable in the face of their mutual bids to change the other's mind.

One tier up, we are arguing that these analyses demonstrate how the choice of cognitive unit can be made empirically by attending to the dynamics and stabilities evident in the data. This is in contrast to the positions on both sides of the ongoing debate about the unit of analysis, with cognitivist and situativists basing their decision largely on theoretical considerations. We introduced four empirical heuristics for identifying the scale of cognitive dynamics. Our analyses of the two episodes highlight the utility and generativity of our empirical heuristics, although we do no claim that these span the set of evidentiary supports for determining the cognitive unit. All of this is underwritten by the resources and framing model of mind, since resources can be distributed within an individual mind or across several individuals and artifacts. Hopefully our approach will enable greater communication and collaboration between cognitivists and those who model cognition as situated and/or distributed.

---

[i] The Newton's 3rd law clip is also discussed in Scherr & Hammer (2009).

[ii] The Shadows and Light clip is also discussed in (Lising & Elby, 2005).